\documentstyle[11pt,aaspp4]{article}
\def\beq{\begin{equation}}
\def\eeq{\end{equation}}
\def\bey{\begin{eqnarray}}
\def\eey{\end{eqnarray}}

\def\sech2{{\rm sech}^{2}}
\input epsf

\begin{document}

\title
	{Extinction bias of microlensed stars towards the LMC
and the fraction of machos in the halo}
\author
	{HongSheng Zhao
	\\Sterrewacht Leiden, 
Niels Bohrweg 2, 2333 CA, Leiden, The Netherlands (hsz@strw.LeidenUniv.nl)}
\date{Accepted ........      Received .......;      in original form .......}
\label{firstpage}

\begin{abstract}
We study the effect of reddening on microlensed stars towards the LMC.
If lenses are in the LMC disk, then the source stars should be at the
far side or behind the LMC disk.  Thus they should experience more
reddening and extinction by dust in the LMC disk than typical stars
in the immediate neighbouring lines of sight.  We simulate
this effect in a variety of models for the LMC stars and dust.  We
stress that the optical depth is not one number, but a function of the
reddening of the survey stars.  We discuss how these effects could be
used to constrain the fraction of machos in the dark halo.  The effect
of patchiness of dust can be controled by working with faint stars in
the smallest patches of the sky around the microlensed stars.
This can be done most effectively with the Hubble Space Telescope in the 
ultra-violet.  The non-detection of the reddening bias would 
be strongly in favor of machos in the Galactic halo.
\end{abstract}


\section{Introduction}

One of the main puzzles of Galactic microlensing surveys is the poorly
determined location of the lens population of the events towards the
Magellanic Clouds.  Currently there are two popular views on the
issue: (a) the lenses are located in the halo, hence are likely
baryonic dark matter candidates (Alcock et al. 1997); (b) both the
lenses and sources are part of the Magellanic Clouds, hence are stars
orbiting in the potential well of the Clouds (Sahu 1994, Wu 1994, Zhao
1998a,b, 1999a, Weinberg 1999).  These lead to drasticly different
predictions the relative amount of baryonic dark matter in galaxy
halos as compared to baryons in stars and gas of various phases in
galaxies and intergalactic medium.  Gould (1995) showed that the low
dispersion of many LMC tracers limits the amount of self-lensing in a
virialized disk of the LMC, as assumed in original models of Sahu and
Wu.  Zhao (1998a) suggested that this constraint can be circumvented
by invoking a plausible amount of unviralized material, particularly
tidally excited material in the immediate surrounding of the LMC disk,
as either foreground lenses or background sources.  Zhao (1998b)
suggested that the polar stream seen in Kunkel et al.'s (1997)
kinematic survey of the LMC carbon stars could contribute to lensing.
Weinberg (1999) showed a specific model for thickening the LMC disk by
interaction with the Galaxy.  Presently the observational status is
still very confusing.  Several observations suggest that our line of
sight to the LMC passes through a 3-dimensional stellar distribution
more extended in the line of sight than a simple thin disc of the LMC
(Kunkel et al. 1997, Zaritsky et al. 1997, 1999 and references
therein).  Zaritsky et al. (1999) also include a review of the
arguments and counter-arguments for these structures.

To break the degeneracy of the models we need new observations which
are sensitive to the thickness of the LMC and the location of the
lenses.  Several ways have been proposed, including studying the
spatial distribution and magnitude distribution of the microlensed
sources (Zhao 1999a, Zhao, Graff \& Guhathakurta 1999), the radial
velocity distribution of the sources (Zhao 1999b) and the reddening
distribution of the sources (Zhao 1999c).  Essentially if lensing is
due to Galactic halo machos or a foreground object, the sample of
lensed stars should be a random subsample of the observed stars in the
LMC.  On the other hand, if there are some tidal material behind the
LMC disk, then the stars in the tidal debris can be strongly lensed by
ordinary stars in the intervening LMC disk, hence the spatial and
kinematical distributions of these source stars will be those of the
tidal debris, rather than those of the LMC disk.  It is suggested that
the source stars in the tidal debris should stand out as outliers in
the radial velocity distribution of the LMC (Zhao 1999b), or
fainter-than-normal red clump stars in the color magnitude diagram
(Zhao, Graff \& Guhathakurta 1999).

The idea about the reddening distribution is to measure the distribution of the
reddening of individual LMC stars in small patchs of sky centered on
the microlensed stars.  Basicly some kind of ``reddening parallaxes''
can be derived for these stars from the line of sight depth effect,
i.e., the dust layer in the LMC makes stars behind the layer
systematicly redder than those in front of the layer.  The method
involves obtaining multi-band photometry and/or spectroscopy of fairly
faint ($19-21$mag) stars during or well after microlensing.  
From these data we can infer the reddening of microlensed stars and
neighbouring unlensed stars.  The method, in some sense, is a variation
of the distance effect suggested by Stanek (1995) for the Galactic bulge
microlensing events.  

Here we extend the analysis of Zhao (1999c) and present simulations of
the reddening of microlensed sources and show how reddening might be
used to constrain the macho fraction in the halo.  The models are
described in \S2, results are shown in \S3.  We summarize in \S4 and
discuss a few practical issues.

\section{Models}

\subsection{Density models for the dust, the stars and the lenses}

For the time being we shall assume a smoothed dust layer of the LMC 
with a $\sech2$ profile and a FWHM of $w$.  
Let $\rho_{\rm d}(D)$ be the dust density at distance $D$, then 
\beq\label{dustden}
\rho_{\rm d}(D) dD= \rho_{\rm d}(D_{\rm LMC}) \sech2 \left({D-D_{\rm LMC} \over 0.56 w}\right) dD,
\eeq
where $\rho_{\rm d}(D_{\rm LMC})$ is
the dust density at the mid-plane of the LMC.

For the LMC stars, we assume a two-component model.
Let $\rho_{\rm LMC}(D)$ be the density of the LMC stars, then
\beq
\rho_{\rm LMC}(D) = {\Sigma_{\rm disk} \over 0.56 W}
\sech2 \left({|D-D_{\rm LMC}| \over 0.56 W}\right) + 
{\Sigma_{\rm extra} \over 1.06 W_{\rm extra}}
\exp\left[{-0.69 L^2}\right],
\eeq
where
\bey
L &=& {(D_{\rm max}-D_{\rm min}) (D-D_{\rm extra})\over  (D_{\rm max}-D_{\rm extra})W_{\rm extra}}, 
~~{\rm if~} D_{\rm max}> D > D_{\rm extra},\\
  &=& {(D_{\rm max}-D_{\rm min}) (D-D_{\rm extra})\over  (D_{\rm extra}-D_{\rm min})W_{\rm extra}}, 
~~{\rm if~} D_{\rm min}< D < D_{\rm extra}.
\eey
Our LMC model includes a (thin) $\sech2$-disk component of a surface
density $\Sigma_{\rm disk}$ and FWHM thickness $W$,
and an extra component with a surface density $
\Sigma_{\rm extra}$ and FWHM thickness $W_{\rm extra}$.  The extra
component consists of two half-Gaussians, which are joined together at their
peaks at $D=D_{\rm extra}$ and truncated at the lower and upper ends at
$D=D_{\rm min}$ and $D=D_{\rm max}$.  This is intended to model any 
non-virialized material in the vicinity of the LMC (Zhao 1998a, Weinberg 1999).
To be most general, an offset between the extra component and 
the LMC disk, namely $D_{\rm extra} \ne D_{\rm LMC}$, is allowed in our models.

Let $\nu_*(D)dD$ 
be the number of LMC star in a line of sight distance bin ($D$, $D+dD$),
then the LMC star number density will be given by
\beq\label{nustar}
\nu_*(D)dD = C_0 \rho_{\rm LMC}(D) D^2 dD,
\eeq
where $C_0$ is a normalization constant.

For the lens population, we include the contribution from the LMC stars
and  intervening machos in an isothermal
halo.  The density of lenses at distance $D_l$ is
\beq\label{lensden}
\rho_{\rm lens} (D_l) = 
\rho_{\rm LMC}(D_l) + f_{\rm macho} \rho_{\rm halo}(D_l),
\eeq
where 
$f_{\rm macho}$ is the fraction of machos in the halo, and
$\rho_{\rm halo}(D)$ is the dark halo density given by
\beq\label{halo}
\rho_{\rm halo}(D) = 0.01M_\odot {\rm pc}^{-3} 
\left[ 1+\left({D \over 8{\rm kpc}}\right)^2 \right]^{-1}.
\eeq
The halo density corresponds to 
a surface density of dark matter (machos plus wimps) halo
$\Sigma_{\rm halo} \sim 100M_\odot\,{\rm pc}^{-2}$ towards the LMC.

\subsection{Reddening vs. line of sight distance of a star 
and their distributions}

In general the reddening of a star is correlated with the distance to
the star, with stars at the backside of the LMC seeing more dust and
experiencing more reddening.  But since the dust distribution is
clumpy and the reddening distribution is patchy, any relation between
the reddening and the line of sight distance is not a well-defined
one-to-one relation, but has a significant amount of scatter due to
patchiness of dust and measurement error.  Nevertheless we can argue
that stars in a small (e.g., $4'' \times 4''$) patch of sky will see
the same set of dust clouds, so if we sort these neighbouring stars
according their reddening, we also get a sorted list in distance.

Stars in the LMC can be ranked according to an observable reddening indicator
$\kappa_{\rm obs}$.  For any LMC star define
\beq\label{kobs}
\kappa_{\rm obs} \equiv
{E(B-V)_{\rm LMC}^{\rm obs} \over \left<E(B-V)_{\rm LMC}^{\rm obs}\right>},
\eeq
where 
$E(B-V)_{\rm LMC}^{\rm obs}$ is the observed reddening in $B-V$ color towards 
the LMC star, after discounting the reddening by the Galactic foreground, and 
$\left<E(B-V)_{\rm LMC}^{\rm obs}\right>$ is the average over random stars
in the small patch of sky centered on that LMC star,
and is approximately the reddening of those stars at the mid-plane of the LMC.
Surely the typical reddening 
$\left<E(B-V)_{\rm LMC}^{\rm obs}\right>$ varies widely from patches
to patches on scales of arcmins (Harris et al. 1997 and references therein), 
but the rescaled reddening 
$\kappa_{\rm obs}$ allows us to sort the stars
not only in the same patch but also stars in different patches.
A higher rank, i.e. a larger $\kappa_{\rm obs}$,
means a deeper penetration into the dust layer.  
Stars at the midplane have a rank
$\kappa_{\rm obs}=1$, and those at the front or back side of the LMC disk
have a rank $\kappa_{\rm obs}=0$ or $\kappa_{\rm obs}=2$.

Consider the reddening rank of stars in a smooth dust model.
Let $A_{\rm LMC}(D)$ be the dust absorption towards a LMC star at distance $D$
(after discounting the absorption by the Galactic foreground dust), 
and let $\kappa(D)$ be the predicted reddening rank of the star, then 
\beq
\kappa(D) = {A_{\rm LMC}(D) \over A_{\rm LMC}(D_{\rm LMC})}=
{\int_0^{D} \!\! \rho_{\rm d}(D) dD \over 
\int_0^{D_{\rm LMC}} \!\! \rho_{\rm d}(D) dD}
\eeq
(cf eq.~\ref{dustden}),
where $\rho_{\rm d}(D_{\rm LMC})$ and $A_{\rm LMC}(D_{\rm LMC})$ are
the dust density and absorption at the mid-plane of the LMC.
It is easy to show that
\bey\label{kd}
\kappa(D) &=& 1+\tanh \left({D-D_{\rm LMC} \over 0.56 w}\right)\\
	  &=& 0,~~~D-D_{\rm LMC} \ll w\\
          &=& 1,~~~D=D_{\rm LMC} \\
	  &=& 2,~~~D-D_{\rm LMC} \gg w.
\eey
So the reddening rank $\kappa(D)$ increases from 0 to 2 abruptly 
in the distance range $D_{\rm LMC} \pm {w \over 2}$ as a star enters and exits
the dust layer.  It turns out that this property is not limited to
$\sech2$ profile dust, but generic for smooth dust layer of a symmetric
profile.

Now to take into account of the scatter due to measurement error and
any residual patchiness, we assume that the observed reddening has
a simple Gaussian distribution centered on the predicted value from a smooth 
dust disk model.  Then for a star at any distance $D$, the reddening rank
$\kappa_{\rm obs}$ is drawn from the following distribution function
\beq\label{broad}
B(\kappa_{\rm obs}, D) = {1 \over \sqrt{2\pi} \sigma} 
\exp \left[-{ \left(\kappa_{\rm obs}-\kappa(D)\right)^2 \over 2 \sigma^2}\right],
\eeq
which is a Gaussian with a constant dispersion $\sigma$, and a mean 
$\kappa(D)$ as predicted from a smooth model.

Now we integrate over LMC stars at all distance and
bin them according to their reddening $\kappa_{\rm obs}$.
Let $N_*(\kappa_{\rm obs})$
be the relative frequencies of finding an unlensed star 
with a reddening $\kappa_{\rm obs}$, then
\beq\label{ns}
N_*(\kappa_{\rm obs}) =  \int dD \nu_*(D)
B(\kappa_{\rm obs}, D).
\eeq

\subsection{Optical depth as a function of distance and reddening}

The microlensing optical depth
$\tau(D_s)$ of a source star at distance $D_s$ is defined by
\beq
\tau(D_s) \equiv {\nu_s(D_s) \over  \nu_*(D_s) }
\eeq
where $\nu_*(D_s)dD_s$ and
$\nu_s(D_s)dD_s$ are the numbers of stars and microlensed sources
in the distance bin ($D_s$, $D_s+dD_s$) respectively (cf. eq.~\ref{nustar}).
It is well-known that in the standard
full-macho isothermal halo model $f_{\rm macho}=1$, 
$\rho_{\rm lens}=\rho_{\rm halo}$ (cf. eq.~\ref{lensden}), and
the microlensing optical depth $\tau_{\rm std}=5\times 10^{-7}$
(cf. e.g., Paczy\'nski 1986).  It is convenient to rescale the optical
depth of a model with this standard value.  The microlensing optical depth
$\tau(D_s)$ is then given by
\beq
\tau(D_s) =\tau_{\rm std} {
\int_{0}^{D_s} \!\! dD_l \rho_{\rm lens}(D_l) {(D_s -D_l)D_l/D_s} 
\over 
\int_{0}^{D_{\rm LMC}} \!\! dD_l 
\rho_{\rm halo}(D_l) {(D_{\rm LMC} -D_l)D_l/D_{\rm LMC}} 
}.
\eeq

The optical depth $\tau(D_s)$ increases with the source distance.
Suppose all lenses are at distance
$D_l$, then
\beq\label{tauprop} 
\tau(D_s) \sim (D_s - D_l).  
\eeq

Unfortunately, it is difficult to observe this effect because it is
hard to measure the distance to a LMC star at the accuracy of 1 kpc.
But fortunately the distance to a star is correlated with the
reddening, and the reddening of a star can be measured accurately to
20\%.  So we have an interesting effect that the rate of microlensing
is correlated with the reddening of the stars.  For each bin in terms
of the rank of reddening ($\kappa_{\rm obs}$, $\kappa_{\rm
obs}+d\kappa_{\rm obs}$), we can define an optical depth
\beq
\tau_{\rm obs}(\kappa_{\rm obs})
\equiv {N_s(\kappa_{\rm obs}) \over N_*(\kappa_{\rm obs})}
={\int dD_s \nu_s(D_s) B(\kappa_{\rm obs}, D_s) 
 \over \int dD \nu_*(D) B(\kappa_{\rm obs}, D)},
\eeq
where
$N_*(\kappa_{\rm obs})$ and $N_s(\kappa_{\rm obs})$
are the relative frequencies of finding an unlensed star and
a microlensed source with reddening $\kappa_{\rm obs}$ respectively
(cf. eq.~\ref{ns}).

Note $\tau_{\rm obs}(\kappa_{\rm obs})$ is a function
measurable observationally because we can always bin the microlensed events
according their observable reddening, and get their frequencies
$N_*(\kappa_{\rm obs})$ and $N_s(\kappa_{\rm obs})$.  Of particular interest is
$\tau_{\rm obs}(0<\kappa_{\rm obs}\le 1)$.  
Note stars at the mid-plane of the LMC disk have
average reddening, so $\kappa_{\rm obs} \sim 1\pm 3\sigma$ at 
$3\sigma$ confidence level.  In comparison stars well in
front of the dust layer have $\kappa_{\rm obs}<3\sigma$, and stars
well behind the dust layer have $\kappa_{\rm obs}>2-3\sigma$.  So 
$\tau_{\rm obs}(0<\kappa_{\rm obs}\le 1)$ is the optical depth of
a thin layer of stars co-spatial with the dust layer and slightly closer to us
the mid-plane of the LMC with $D_{\rm LMC}-w/2 \le D \le D_{\rm LMC}$.  
Since the thickness of the dust layer $w \sim 100-200$pc is very small,
these stars are virtually at the same distance.
So we have
\beq\label{tobs1}
\tau_{\rm obs}(0 <\kappa_{\rm obs}\le 1) 
\approx \tau(D_{\rm LMC}) \ge 
f_{\rm macho}\tau_{\rm std}.
\eeq
So we have turned the theoretical quantity $\tau(D_{\rm LMC})$ to
an observable, and this observable optical depth 
$\tau_{\rm obs}(0<\kappa_{\rm obs}\le 1)$ 
sets an upper limit on the amount of machos in the foreground of the LMC.
The above inequality reduces an equality 
in the absence of any LMC lenses in front of the LMC disk.

If the picture that the LMC is a thin disk without any extra material
in the front or back is correct, then we should find that $\tau_{\rm
obs}(\kappa_{\rm obs})$ is at a constant level 
$\approx \tau(D_{\rm LMC})= f_{\rm macho}\tau_{\rm std}$
for all reddening.

Presently the survey teams have not studied the reddening dependence
of the optical depth, so they quote only the observed optical depth
averaged over all LMC stars, which is only one number.
Let this be $\left< \tau_{\rm obs} \right>$, then current data require
\beq\label{tobs}
\left< \tau_{\rm obs} \right> \equiv
{\int_{D_{\rm min}}^{D_{\rm max}}\!\! dD \nu_*(D) \tau(D) 
\over \int_{D_{\rm min}}^{D_{\rm max}}\!\! dD \nu_*(D) } 
={\int\!\! d\kappa_{\rm obs} N_*(\kappa_{\rm obs}) \tau(\kappa_{\rm obs}) 
\over \int\!\! d\kappa_{\rm obs} N_*(\kappa_{\rm obs}) } 
\approx 0.5 \tau_{\rm std},
\eeq
where the factor $0.5$ comes from the fact 
that the observed microlensing optical
depth account for about one-half of that of the standard macho dark halo model 
$\tau_{\rm std}$ (Alcock et al. 1997).

We would like to know how much of the observed optical depth 
is due to stellar lenses in the LMC vicinity and how much
is due to machos in Galactic halo.
If we define $f_*$ as the fraction of the observed
microlensing depth due to stellar lenses, then the fraction of
machos in the dark halo is given by
\beq\label{fstar}
f_{\rm macho} = {\left< \tau_{\rm obs} \right> (1-f_*) 
\over \tau_{\rm std}} = 0.5 (1-f_*).
\eeq
So for a given set of stellar distribution parameters $\Sigma_{\rm
extra}$ etc, we can predict the optical depth due to stellar lenses
$f_*\tau_{\rm std}$, we can then determine
$f_{\rm macho}$ by fixing the overall optical depth to the 
observed value.

Finally we would like to characterize
the excess reddening of the microlensed stars.  
This can be done by defining an observable parameter
$\xi_{\rm obs}$ such that
\bey
\xi_{\rm obs} &\equiv &
{\int \! \! d\kappa_{\rm obs} N_s(\kappa_{\rm obs}) \kappa_{\rm obs}\over
 \int \! \! d\kappa_{\rm obs} N_s(\kappa_{\rm obs}) }-
{\int \! \! d\kappa_{\rm obs} N_*(\kappa_{\rm obs}) \kappa_{\rm obs}\over
 \int \! \! d\kappa_{\rm obs} N_*(\kappa_{\rm obs}) }\\
&=&
{\int \! \! dD \nu_s(D) \kappa(D) \over
 \int \! \! dD \nu_s(D)  }-
{\int \! \! dD \nu_*(D) \kappa(D) \over
 \int \! \! dD \nu_*(D)  }.
\eey
Because of the integration over the entire range of the reddening, 
$\xi_{\rm obs}$ is independent of the dispersion $\sigma$ 
between the observed reddening rank of a star and the reddening rank 
predicted from a smooth model
(cf. eq.~\ref{broad}).  

\section{Results}

\subsection{Model parameters}

Here we study the effect of a mixed halo model on the reddening
distribution of the microlensed stars.  Since the parameters of the
LMC are very uncertain, we will explore a range of dust models and
stellar models.  Table 1 lists the parameters for the models.  

We fix the distance to the LMC $D_{\rm LMC}=50$kpc, and set the tidal
radius of the LMC $|D_{\rm min}-D_{\rm LMC}|=|D_{\rm max}-D_{\rm
LMC}|=10$kpc.  For the $\sech2$ stellar disk of the LMC and the dust
layer, we fix $\Sigma_{\rm disk}=400M_\odot{\ \rm pc}^{-2}$.  We fix
the FWHM thickness of the dust layer $w=200$ pc, but allow the
relative thickness of the two disks $W/w$ to vary in the range 0.5 to
2, a reasonable range for thin stellar and dust disks.  We set the
FWHM of the tidal material $W_{\rm extra}=0.05(D_{\rm max}-D_{\rm
min})=1$kpc, but allow the peak position to vary in the range 40 kpc
$\le D_{\rm extra} \le$ 60 kpc, and the amount of stars in extra
component to vary in the range $0 \le \Sigma_{\rm extra} \le
50M_\odot\,{\rm pc}^{-2}$.  We set this upper limit for the surface
density of the extra material to about $10\%$ of the surface density
of the LMC, an acceptable amount for some hidden material.  In
comparison the surface density of dark matter (machos plus wimps) halo
$\Sigma_{\rm halo} \sim 100M_\odot\,{\rm pc}^{-2}$ (cf. eq.~\ref{halo}).
These parameters are comparable to those of previous models for the volume
density of the LMC disk (Wu 1995, Weinberg 1999).

It turns out the reddening distributions of microlensed and unlensed
stars are insensitive to exact values of $w$, $W$, $W_{\rm extra}$ and
$\Sigma_{\rm disk}$ as long as we are in the regime where $w \sim W
\ll W_{\rm extra} \ll |D_{\rm max}-D_{\rm min}|$ and $\Sigma_{\rm
extra} \sim \Sigma_{\rm halo} \ll \Sigma_{\rm disk}$.  
These distributions are more sensitive
to the ratios such as $W/w$, $\Sigma_{\rm extra}/\Sigma_{\rm halo}$,
and $|D_{\rm extra}-D_{\rm LMC}| \over |D_{\rm max}-D_{\rm min}|$.
For example, models with the extra stars centered on the LMC disk,
i.e., ${|D_{\rm extra}-D_{\rm LMC}| \over |D_{\rm max}-D_{\rm min}|}
\sim 0$, require much more stars with $\Sigma_{\rm extra}/\Sigma_{\rm
halo} \gg 1$ for come up with enough events, hence are much less
efficient than models with extra stars $5-10$kpc in front or behind
the LMC in terms of producing microlensing.
So the main parameters that we will vary are
$\Sigma_{\rm extra}$, $D_{\rm extra}$ and $W/w$.
The predictions are made for halo-lensing models and mixed models with
different amount of stars in the extra component and different offset
distance from the LMC and the relative thickness of the stellar disk
vs. the dust disk.

We also set the random error $\sigma$ for the observed reddening
at 20\%-40\% level, which seems a reasonably amount.  A larger
dispersion would smooth out some tell-tale features of the reddening
distributions.  But it turns out that the mean excess $\xi_{\rm obs}$
and and the optical depth in the mid-plane $\tau_{\rm obs}(\kappa_{\rm
obs}=1)$ are insensitive to the observational error $\sigma$.

\subsection{Reddening distributions of microlensed and unlensed stars}

\begin{figure}
\epsfxsize=10cm \centerline{\epsfbox{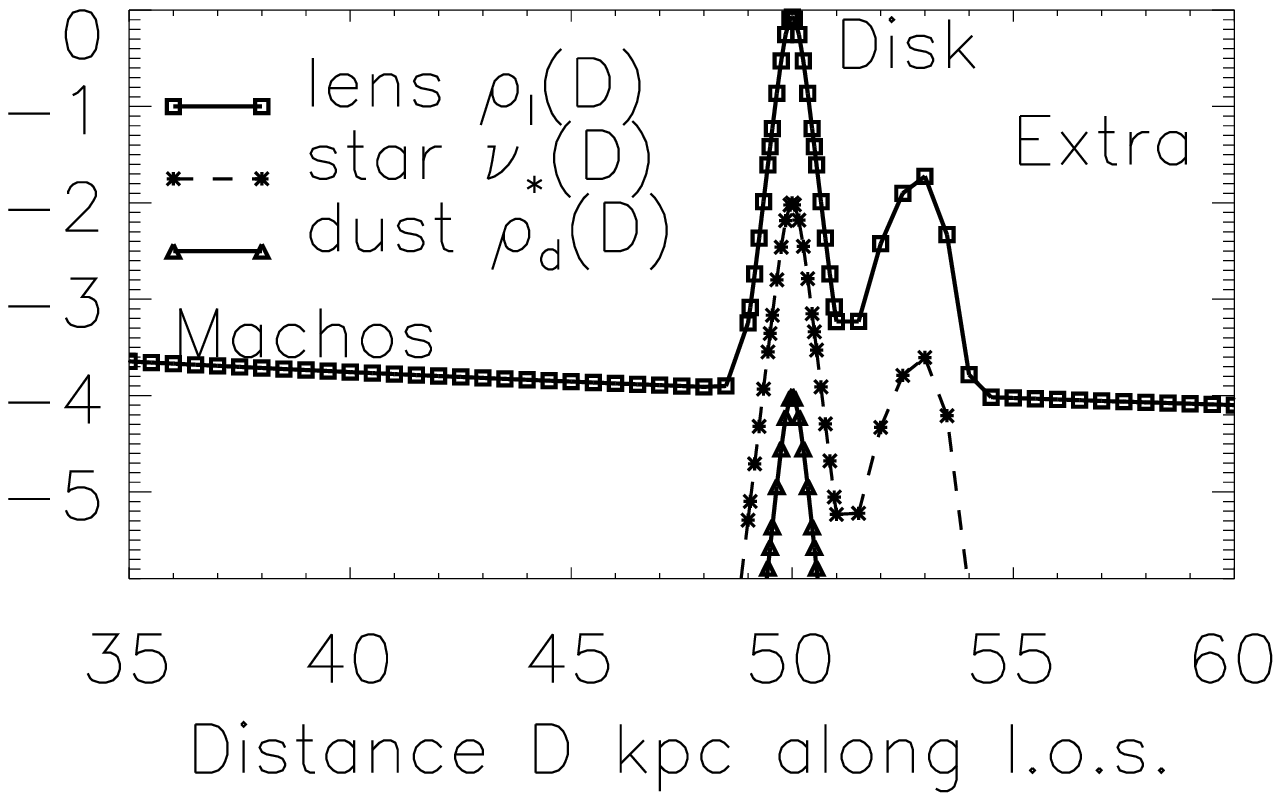}}
\epsfxsize=10cm \centerline{\epsfbox{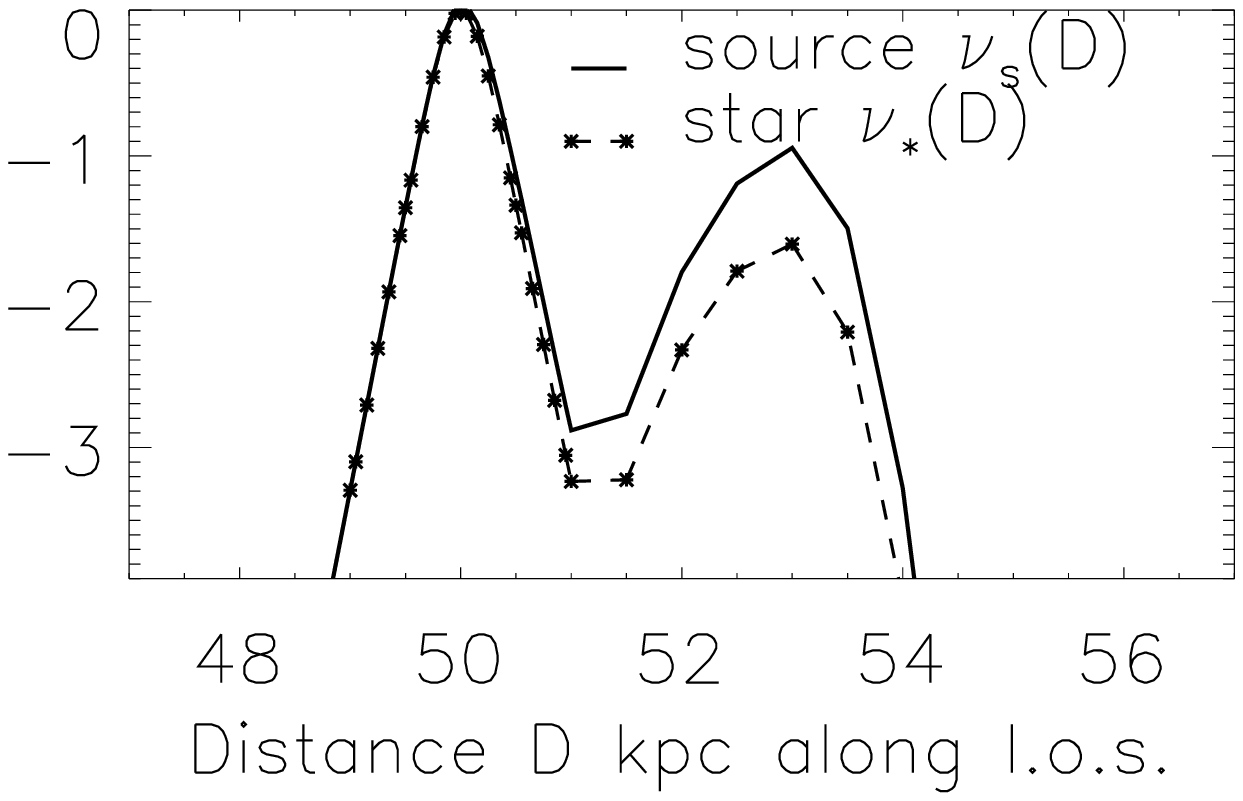}}
\epsfxsize=10cm \centerline{\epsfbox{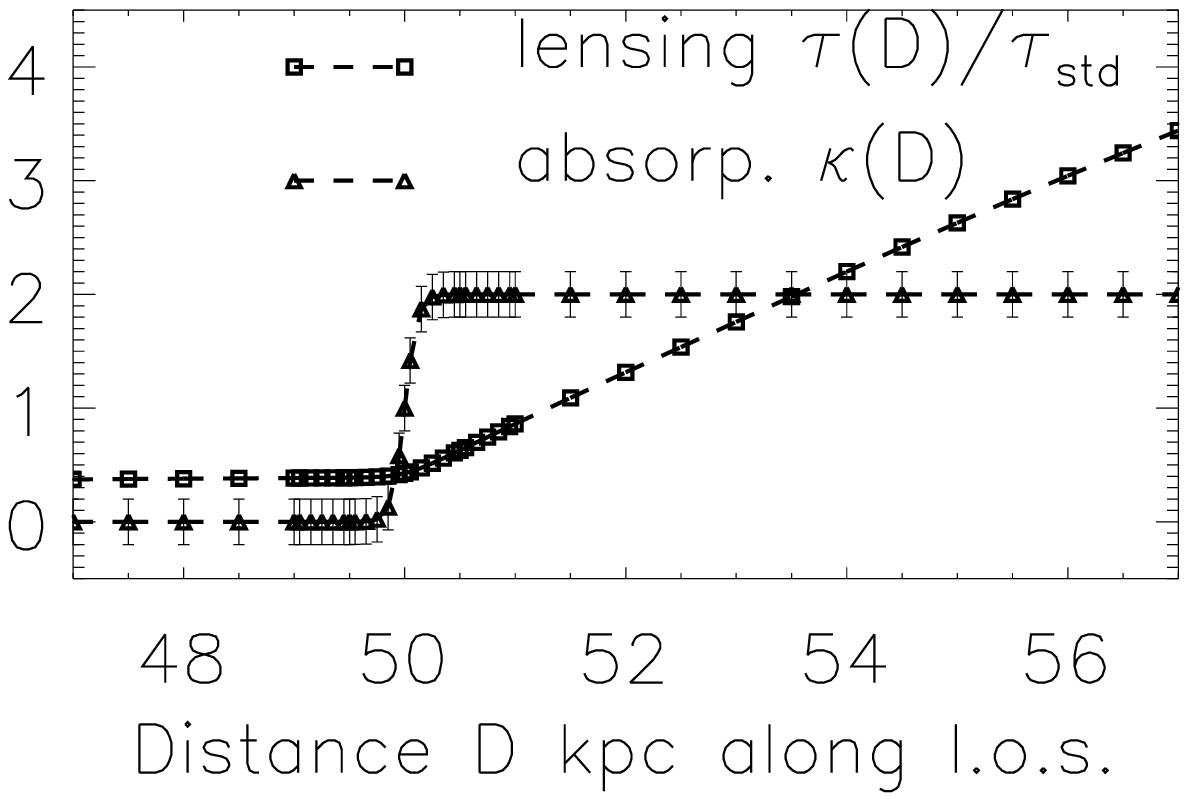}}
\caption{
Upper panel: the density (in logarithmic scale with arbitrary zero point)
of the lenses, the unlensed LMC stars and
the density of the dust layer along the line of sight to the LMC.
Middle panel: a zoom-in of the upper panel for the unlensed LMC stars
and the lensed source stars.
Lower panel: the run of the rescaled
absorption of the dust (dashed lines with error bars) 
and the rescaled optical depth of microlensing (dashed line with squares) 
as functions of line of sight distance.  The small error bars 
indicate a 20\% dispersion due to patchiness and measurement error.
All calculations are done for $W=2w=400$pc, $D_{\rm extra}=53$kpc, 
$\Sigma_{\rm extra}=20M_\odot\,{\rm pc}^{-2}$ and $f_{\rm macho}=0.38$.
}
\label{den.ps}
\end{figure}

The line of sight distributions of a few model quantities are shown in
Fig.~\ref{den.ps} for one of the models (see caption).  Clearly all
stellar density distributions peak at $D=D_{\rm LMC}=50$kpc, the
mid-plane of the high density disk of the LMC; similarly for the thin
dust layer.  The LMC stars have a second peak due to extra stars
placed behind the LMC disk.  The lens density distribution also has a
gentle falling part in the range 10 kpc $<D<$ 40 kpc due to foreground
machos, whose number density falls as $D^{-2}$.  The lensed stars
(dotted line with diamonds) also have a higher tail.

For sources in front the LMC disk $D<D_{\rm LMC}$ the microlensing
optical depth is nearly constant, $\tau(D_s)/\tau_{\rm std} \sim f_{\rm macho} 
\sim 0.36$.  This is because the optical depth due to
the ever decreasing density of the foreground machos is insensitive to
the source distance.  But after passing the LMC disk the optical depth
takes off linearly with the source distance till $\tau(D_s)/\tau_{\rm std}
\gg 1$.  This is because any source star behind the LMC suddenly sees
an intervening dense screen of lenses coming from the high surface
density stellar disk of the LMC, so the optical depth goes up
in proportional to the distance to the LMC disk (cf. eq.~\ref{tauprop}). 
This explains 
the higher tail of the lensed stars as compared to unlensed stars 
(cf. Fig.~\ref{den.ps}).  Note that there is only a modest increase
of the optical depth from entering to exiting the LMC thin disk.
The amount is on the order of 
$(\Sigma_{\rm disk}/\Sigma_{\rm halo})(W/10{\rm kpc})<8\%$ of the total
optical depth, consistent with
the estimation of Sahu (1994) and Wu (1994) for a thin disk.

The dust extinction is nearly a Heaviside function of the distance,
climbing steeply from zero absorption in front of the layer to a
constant value $2\times A_{\rm LMC}(D_{\rm LMC})$ a few hundred pc
behind the layer, where $A_{\rm LMC}(D_{\rm LMC})$ is the absorption
in the mid-plane.  This is because as a star sinks deeper in the dust
layer it experiences an increasing reddening before it emerges from the
dust layer again.

Fig.~\ref{redist.ps} shows the reddening distributions of microlensed
and unlensed stars.  The distribution for unlensed stars (upper
panels) hardly depends on parameters of the extra component
$\Sigma_{\rm extra}$ and $D_{\rm extra}$.  It is only sensitive to the
relative thickness of the stellar disk and the dust layer.  At one end
of the extreme, we have a thin dust layer and a thicker stellar disk
($W/w=2$, solid lines).  We see the familiar double-horn structure.
This is because most stars in a thick disk are either in front of the
dust layer and free from reddening or behind the layer and reddened by
the maximum amount.  At the opposite extreme, if the stellar disk is
thinner than the dust layer ($W/w=0.5$, dash-dot-dot-dot lines), the
distribution becomes peaky in the middle because most stars of the
thin disk are in the mid-plane of the dust layer, hence reddened by
half of the maximum (back-to-front) value.  In between if the dust
layer and the stellar disk have the same scale height and are evenly
mixed ($W/w=1$, dashed lines), we have a flat top-hat distribution.

The distributions for the microlensed stars depend on the location and
the amount of the extra component.  For halo-lensing models, the
distribution of lensed stars and unlensed stars are virtually
indistinguishable (cf. Fig.~\ref{redist.ps}a).  The situation is
largely unchanged if we move all lenses from the halo to 10 kpc in
front of the LMC disk (cf. Fig.~\ref{redist.ps}b).  The extra
component shows up merely as a marginal excess of low reddening stars in the
unlensed stars, a small effect which might escape detection.

The situation is drasticly different if we put even a small amount
($\Sigma_{\rm extra}=10M_\odot\,{\rm pc}^{-2}$, or 2.5\%) of stars behind the
LMC disk (Fig.~\ref{redist.ps}c).  The distribution are strongly
skewed towards high reddening.  This is because lensing favors source
stars well-behind the lenses.  More specificly if the lenses are
machos midway to the LMC, or stars some 10 kpc in front of the LMC,
then the probability of drawing a source star at 100 pc behind the
dust layer is about 0.8\% or 2\% higher than drawing one at 100 pc in
front of the dust layer (cf. eq.~\ref{tauprop}).  So in average, the
lensed sources should be more reddened than the unlensed ones by
merely a few percent.  In contrast, if the lenses are at mid-plane of
the LMC, then nearly all sources will be behind the dust layer, and
hence the reddening rank $\kappa_{\rm obs}$ increases by one
(cf. eq.~\ref{kobs} and eq.~\ref{kd}).

In Fig.~\ref{redist.ps}c there are still some source stars at low
reddening because we allow for machos in the foreground.  The skewness
effect becomes stronger as we decrease the macho fraction, and
increase the extra stars behind the LMC.  For $\Sigma_{\rm
extra}=40M_\odot\,{\rm pc}^{-2}$ and $D_{\rm extra}=60$kpc, then all
lensed sources should have an observed reddening rank $\kappa_{\rm
obs} \approx 2$.  In this case $f_{\rm macho} \approx 0$ and all
lenses come from the LMC disk.

Another way to quantify the above effects is to look at the observed
optical depth as a function of the reddening rank of stars
(cf. Fig.~\ref{opt.ps}a).  The plateau of the $\tau_{\rm
obs}(\kappa_{\rm obs})$ curves in the range $0<\kappa_{\rm obs}\le 1$
are for stars imbeded in the dust layer of the LMC.  They are slightly
closer to us than the stars at the mid-plane, for which $\kappa_{\rm
obs}=1\pm 3\sigma$, but are virtually at the same distances $D_{\rm
LMC} \pm w/2$.  The plateau also set an upper limit on the optical
depth of machos (cf. eq.~\ref{tobs1}).

On the other hand those stars with $\kappa_{\rm obs} \ge 2$ include
both stars well behind the LMC disk, which have a very high optical
depth (cf. the curve of $\tau(D)$ in Fig.~\ref{den.ps}), and any
Gaussian tail of the LMC disk stars due to random error of the
reddening.  So $\tau_{\rm obs}(\kappa_{\rm obs})$ can have a high tail
if the extra component behind the LMC is stronger than the Gaussian
tail due to a non-zero dispersion $\sigma$.  These results depend
somewhat on the reddening and the relative thickness of the stellar
disk and the dust layer (cf. Fig.~\ref{opt.ps}b).  A larger
measurement error smoothes the sharp transition from the plateau to
the high tail.  The tail is lower if the stellar disk is thicker than
the dust layer, because of more overlap in the observed reddening of
the stars in the LMC disk and those well-behind the LMC.

Fig.~\ref{fmacho.ps} shows that in general a strong excess in
reddening is seen in models with most of the lenses in the LMC disk,
and the sources behind the midplane of the LMC disk.  Models with a
high macho fraction $f_{\rm macho} \ge 0.4$ and/or with lenses being
extra stars a few kpc well in front of the LMC disk predict only a
modest amount of excess $\xi_{\rm obs} \le 0.1$.  So these models
could be ruled out if we measure a strong excess.  In particular, if
$\xi_{\rm obs} \ge 0.5$ then the dark halo cannot have more than 15\%
in machos.  On the other hand, if the excess is small, then the
interpretation remains non-unique.  These results depends only
slightly on the thickness of the dust layer with the thinner dust
layer predicting a stronger excess.  Given accurate reddening, it is
possible to measure or at least set an upper limit on the fraction of
dark matter in machos.

\subsection{Empirical relations}

We have also find a few empirical relations for converting the 
reddening excess $\xi_{\rm obs}$ to the macho fraction $f_{\rm macho}$.
Assuming that there
are few stellar lenses in the immediate foreground of the LMC disk, 
which would be the case if $0.8 \ge \xi_{\rm obs} \ge 0.2$, then to a good 
approximation, the fraction $f_*$ of microlensing due to LMC stars is given by
\beq
f_* \approx 1.2\xi_{\rm obs}.
\eeq
The fraction of machos is related to $\xi_{\rm obs}$ by
\beq
f_{\rm macho} \approx 0.5(1-1.2\xi_{\rm obs})
\eeq
(cf. eq.~\ref{fstar} and ~\ref{tobs}).
Here we have adopted the present reported value 
for the microlensing optical depth
to the LMC, i.e., about half of that of the standard halo.
But the results can be rescaled proportionally as more data become available.

The result can also be recast as following,
\beq
f_{\rm macho} 
\approx {\tau_{\rm obs}(0<\kappa_{\rm obs}\le 1) 
\over 5\times 10^{-7} }
\eeq
(cf. eq.~\ref{tobs1}), where
$\tau_{\rm obs}(0<\kappa_{\rm obs}\le 1)$ is the optical depth of
the LMC stars observed with lower than average reddening; these are
stars sandwiched in a thin layer 
between the mid-plane of the LMC disk and the near side of the dust layer.
The optical depth of the least reddened stars best approximates
the optical depth of foreground machos, because it is least contaminated
by stellar lenses in the LMC.

These relations work best if we can neglect stellar lenses in the
immediate foreground of the LMC disk.  But interestingly, these results
are insensitive to the assumption of the dust distribution and stellar
distribution of the LMC, e.g., the distance of the extra background
material, and their surface density, the relative thickness of the
dust layer and the stellar disk.  We also allow for a realistic amount
of measurement error and patchiness of extinction.

\section{Summary}

In summary, we have studied effects of dust layer in the LMC on the
microlensing events in a wide range of models of star and dust
distributions in the LMC.  We propose to bin LMC stars according to
their reddening rank $\kappa_{\rm obs}$ (as defined in
eq.~\ref{kobs}), and study the observable microlensing optical depth
as a function of the reddening rank $\kappa_{\rm obs}$.  We find that
self-lensing models of the LMC draw preferentially sources behind the
dust layer of the LMC, and hence can be distinguished from the
macho-lensing models once the reddening by dust is measured.  

\begin{figure}
\ \\ \epsfxsize=0.38\hsize\epsfbox{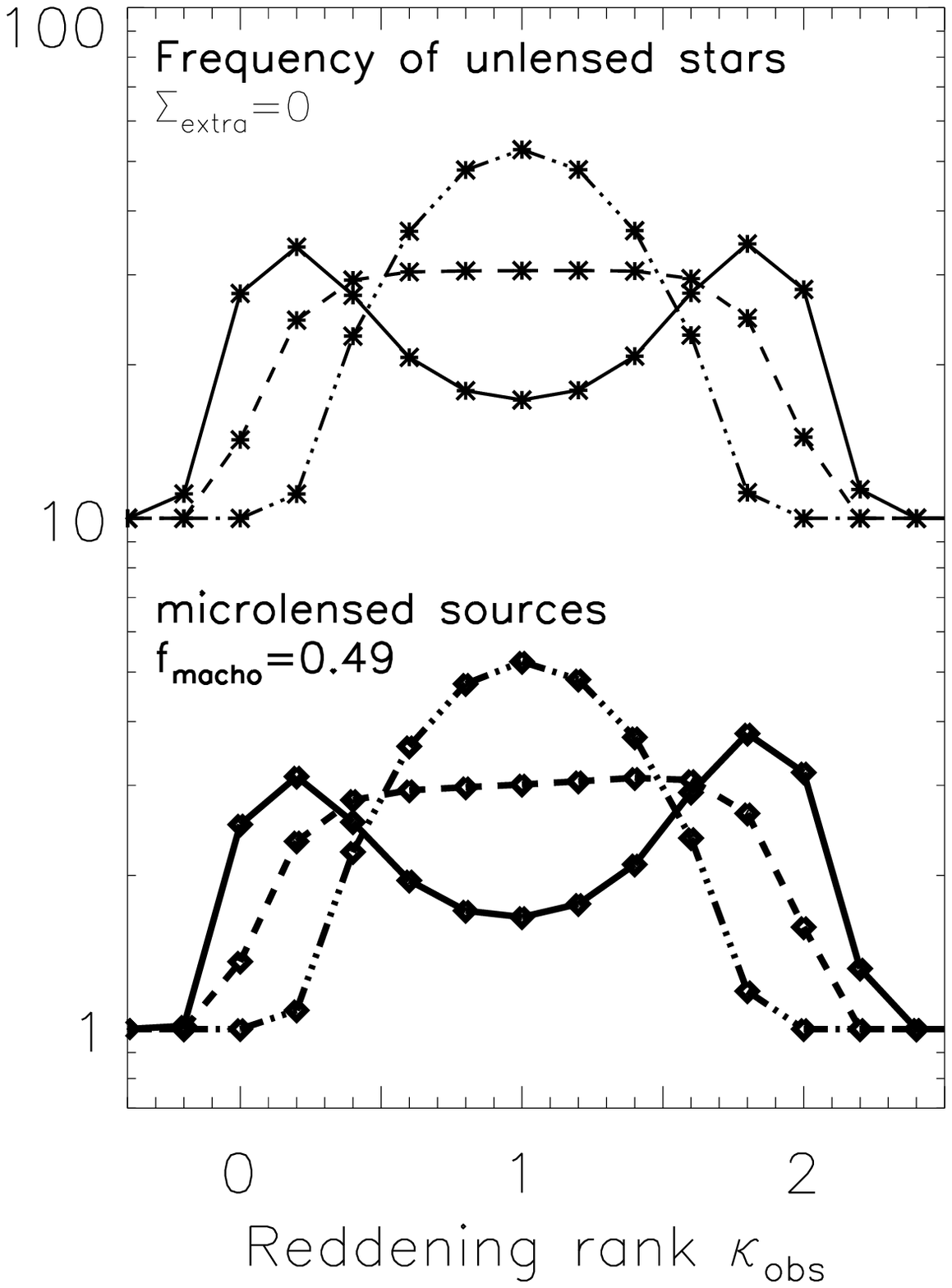}\hspace{-0.09\hsize} \epsfxsize=0.38\hsize\epsfbox{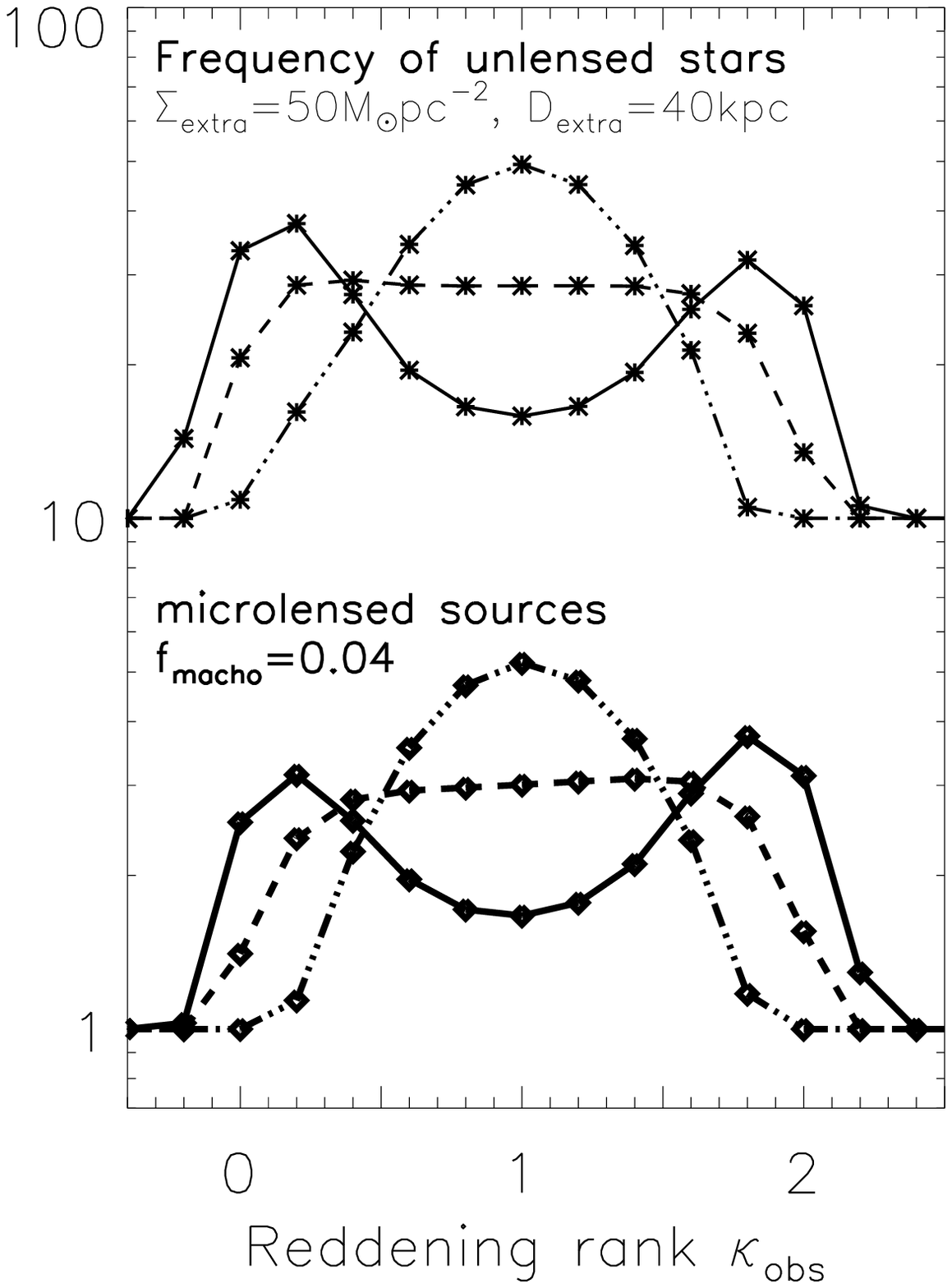}\hspace{-0.09\hsize} \epsfxsize=0.38\hsize\epsfbox{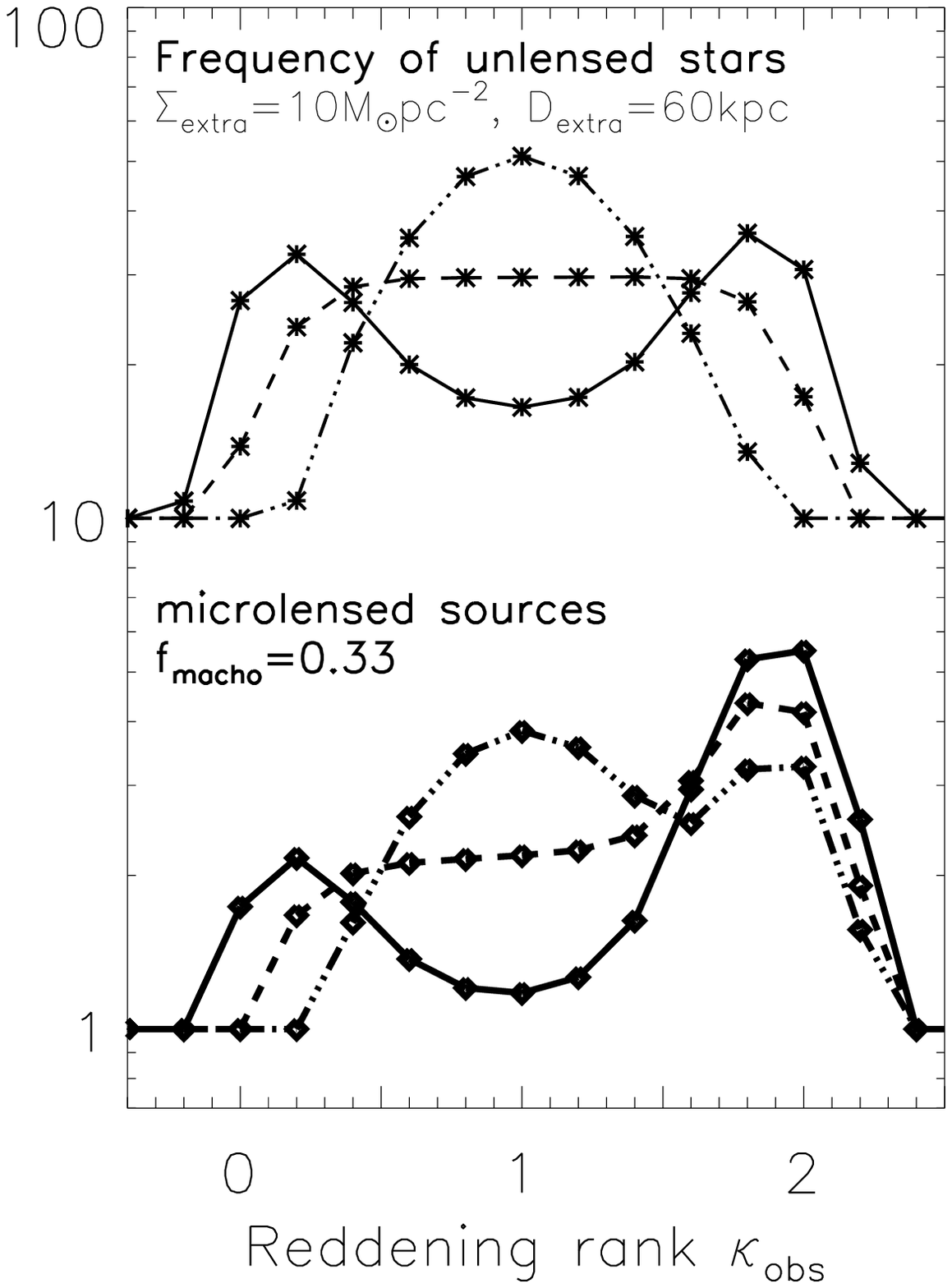}\hspace{-0.09\hsize}
\caption{
the frequency of finding stars with a certain rescaled reddening for the
microlensed sources (the lower part of each panel) and the unlensed
stars (the upper part of each panel).  The legends show model parameters.
Models are calculated for
a thick disk and a thin dust layer with $W=2w=400$pc (solid lines),
an equally thin disk and dust layer with $W=w=200$pc (dashed lines), and
a thin disk and a thick dust layer with $W=w/2=100$pc (dash-dot-dot-dot lines).
The distributions have been smoothed by a Gaussian with a dispersion 
$\sigma=20\%$ to allow for the measurement error and patchiness.  
}
\label{redist.ps}
\end{figure}

If a low excess in reddening ($\xi_{\rm obs}<0.2$) is observed then the
interpretation would not be unique: the reddening distribution does
not distinguish very well lensing by machos and lensing by stars a few
kpc in front of the LMC.  The optical depth can be explained equally
well by a screen of macho lenses with $\Sigma_{\rm extra}=50M_\odot\,{\rm
pc}^{-2}$ at 10 kpc from the Sun, or by a screen of stellar lenses
with a same surface density at 10 kpc in front of the LMC disk.  As
far as the reddening distributions are concerned, these two models are
barely distinguishable (cf. Fig.~\ref{redist.ps}ab).  On the other
hand the reddening distribution is very sensitive to any stars behind
the LMC disk.  If 2-3\% of the LMC stars is put in an extra component $5-10$
kpc behind the LMC, then we can see markedly different distribution in
the microlensed stars (cf. Fig.~\ref{redist.ps}c); this corresponds to
a surface density $\Sigma_{\rm extra}=10M_\odot\,{\rm pc}^{-2}$.  The
signal gets even stronger as we decrease the macho fraction and
increase the surface density of the background stars.

\begin{figure}
\ \\ \epsfxsize=0.5\hsize\epsfbox{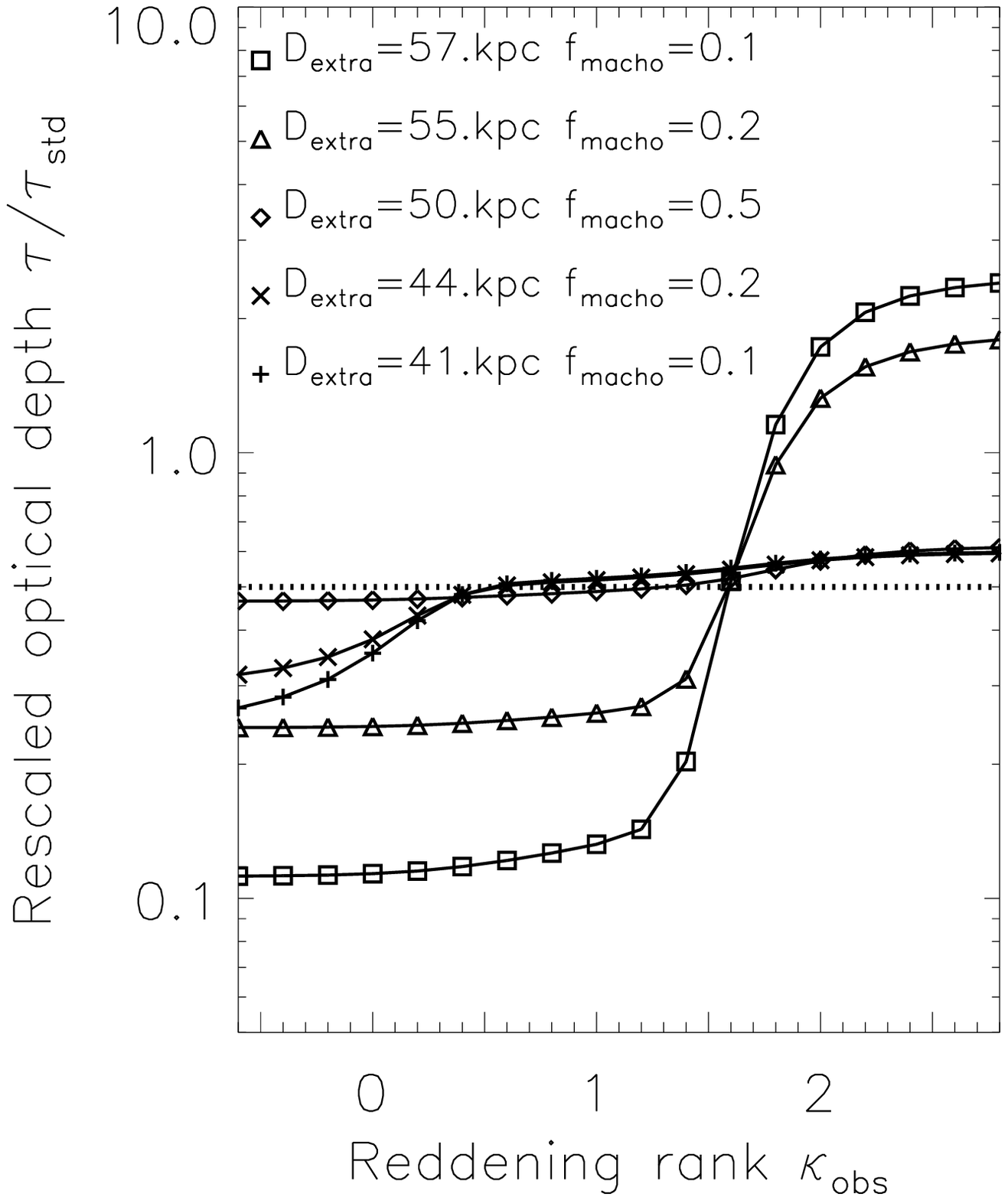}\hspace{-0.03\hsize} \epsfxsize=0.5\hsize\epsfbox{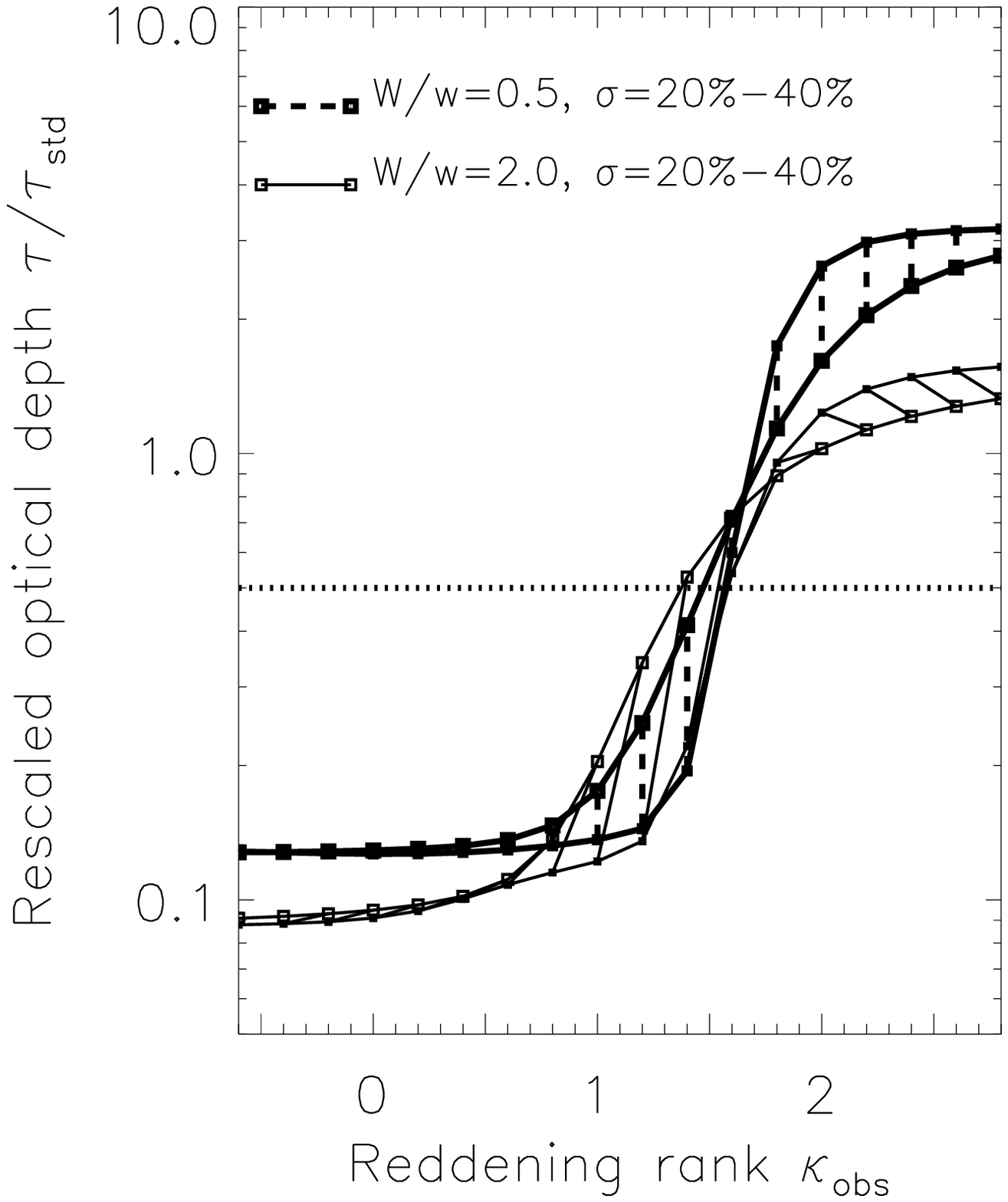}\hspace{-0.03\hsize}
\caption{
the optical depth, rescaled by that of the standard model $\tau_{\rm std}$,
as a function of the reddening for several models with
$\Sigma_{\rm extra}=50M_\odot\,{\rm pc}^{-2}$.
Models are calculated for $\sigma=20\%$, $W=w=200$pc, 
and a range of $D_{\rm extra}=41-57$ kpc (left panel), and for 
$D_{\rm extra}=57$ kpc, and a range of 
$\sigma=20\%-40\%$ and $W/w=0.5-2$ (right panel).
In the absence of stellar lenses, a halo model would predict
the (horizontal dotted) line $\tau_{\rm obs}/\tau_{\rm std}=0.5$. 
}
\label{opt.ps}
\end{figure}

\begin{figure}
\epsfxsize=8cm \centerline{\epsfbox{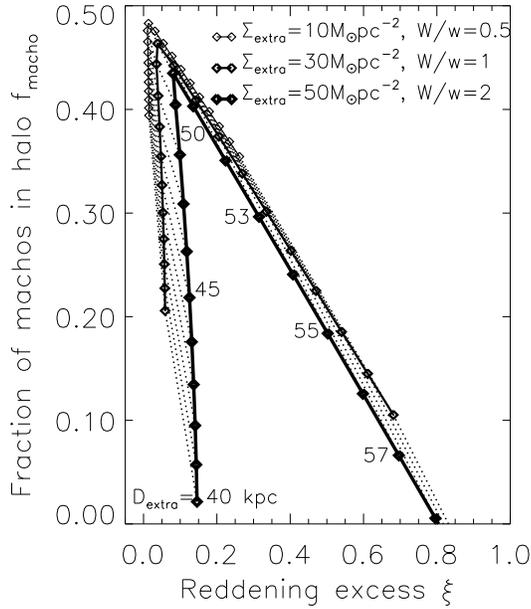}}
\caption{
the excess reddening vs. the fraction of machos.
We vary surface density
$\Sigma_{\rm extra}=(10-50)M_\odot\,{\rm pc}^{-2}$
of the extra stars and its peak position $D_{\rm extra}=40-60$ kpc.  
Each model allows a range of thickness for the dust layer $w$
stellar disk and $W$.  We trade between the amount of extra stars
$\Sigma_{\rm extra}$ and the amount of machos $f_{\rm macho}$ such that
models all have a fixed optical depth 
$\left<\tau_{\rm obs}\right>/\tau_{\rm std}=0.5$.  The
predictions are shown as three hatched bands.}
\label{fmacho.ps}
\end{figure}

Our main finding is that among stars of different reddening rank, 
the optical depth of the least reddened stars
$\tau_{\rm obs}(\kappa_{\rm obs} \sim 0)$ is the closest approximation to
the optical depth of machos (cf. Fig.~\ref{opt.ps}).
The macho fraction $f_{\rm macho}$ also correlates
tightly with the excess reddening $\xi_{\rm obs}$ of the microlensed sources
(cf. Fig.~\ref{fmacho.ps}).  
These results are summarized by the
empirical relations in the previous section.  An observable high
excess in reddening ($1> \xi_{\rm obs} >0.2$) would be a definative signal for
many LMC disk lenses and low macho fraction.  Potentially we can
constrain the fraction of Galactic dark matter in machos this way.

There are a number of problems to apply the method to observations.
(a) Reddening of individual stars is difficult to measure accurately.
Reddening can be determined by constructing reddening-free indices
with photometry of three or more broad bands, or with low resolution
spectroscopy. Typical accuracy is about $0.02$ mag. in $E(B-V)$
with these methods (Harris et al. 1997), 
which is about 20\% of the typical reddening in the LMC (Harris et al. 1997).
(b) Stars in the LMC are likely crowded from the ground.
Blended images of stars lead to unphysical colors and spurious
reddening.  (c) Dust distribution is very clumpy.  Extinction towards
OB stars in the LMC disk can
easily vary at a factor of two level among different patches of the
sky separated by $1'$ (Harris et al. 1997).  
(d) The effect of extinction by the dust layer of the
Galaxy near the Sun should be included.  
Oestriecher et al. (1996) show that the Galactic foreground extinction
is not entirely smooth.  There are dark patches on $30'$ scales.  

The clumpiness of the dust, together with the fairly large error of the
reddening vector measurable from broad-band photometry, can lead to a
large scatter in the relation between reddening and line of sight
depth.  Zhao (1999c) argue that these problems, particularly
patchiness can be overcome if we restrict to the smallest patches of
sky around each microlensing source where the variation is likely at
only 20\% level.  Spectroscopic observation with the Hubble Space
Telescope in the ultra-violet can in principle resolve many faint
stars in the small patches around each microlensed stars, and measure
the reddening accurately.  By obtaining the reddening of stars in the present
$20-30$ microlensing lines of sight we
should be able to measure the excess reddening confidently.
We should also be able to bin the events in the reddening rank, and study
the optical depth as a function of reddening.  A null result of any
variation of the optical depth would be in favor of 
a significant baryonic dark component of the Galaxy.

\vskip 1cm


\vfill \eject

\begin{deluxetable}{lrrrrrrrr}
\tablewidth{0pc}
\tablecaption{Fixed parameters for the stellar, dust and halo models}
\tablehead{
 \colhead{$\Sigma_{\rm disk}$}
& \colhead{$W_{\rm extra}$}
& \colhead{$D_{\rm min}$}
& \colhead{$D_{\rm max}$}
& \colhead{$w$}
& \colhead{$\left<\tau_{\rm obs}\right>$}
& \colhead{$\tau_{\rm std}$}
& \colhead{$\Sigma_{\rm halo}$}
}
\startdata
400$M_\odot{\rm pc}^{-2}$ & 1kpc & 40kpc & 60kpc & 200pc & $2.9\times 10^{-7}$ & $4.7\times 10^{-7}$ & 200$M_\odot{\rm pc}^{-2}$ \nl
\enddata
\end{deluxetable}

\begin{deluxetable}{lrrrr}
\tablewidth{0pc}
\tablecaption{Variable parameters for the stellar, dust and halo models}
\tablehead{
 \colhead{$\Sigma_{\rm extra}$}
& \colhead{$D_{\rm extra}$}
& \colhead{$W/w$}
& \colhead{$\sigma$}
& \colhead{$f_{\rm macho}$}
}
\startdata
0-50$M_\odot{\rm pc}^{-2}$ & 40-60 kpc & 0.5-2 & 20\%-40\% & 0-0.5 \nl
\enddata
\end{deluxetable}

\vfill \eject

{}

\onecolumn

\label{lastpage}

\end{document}